\documentclass[aps,prl,twocolumn,superscriptaddress,showpacs,nobibnotes]{revtex4-1}
\usepackage{graphicx,amsmath,amssymb,ulem,color,braket}
\usepackage{float}

\begin{document}

\title{Exciton-polariton X-waves in a microcavity}

\author{Oksana Voronych}
\address{Institute of Theoretical Physics and Astrophysics, University of Gda\'nsk, ul.~Wita Stwosza 57, 80-952 Gda\'nsk, Poland}

\author{Adam Buraczewski}
\address{Institute of Theoretical Physics and Astrophysics, University of Gda\'nsk, ul.~Wita Stwosza 57, 80-952 Gda\'nsk, Poland}

\author{Micha\l\, Matuszewski}
\address{Institute of Physics, Polish Academy of Sciences, Al.~Lotnik\'ow 32/46, 02-668 Warsaw, Poland}

\author{Magdalena Stobi\'nska}
\email{magdalena.stobinska@gmail.com}
\thanks{Corresponding author.}
\address{Institute of Theoretical Physics and Astrophysics, University of Gda\'nsk, ul.~Wita Stwosza 57, 80-952 Gda\'nsk, Poland}
\address{Institute of Physics, Polish Academy of Sciences, Al.~Lotnik\'ow 32/46, 02-668 Warsaw, Poland}

\begin{abstract}
We investigate the possibility of creating X-waves, or localized wave packets, in resonantly excited exciton-polariton superfluids. We demonstrate the existence of X-wave traveling solutions in the coupled exciton-photon system past the inflection point, where the effective mass of lower polaritons is negative in the direction perpendicular to the wavevector of the pumping beam. Contrary to the case of bright solitons, X-waves do not require nonlinearity for sustaining their shape. Nevertheless, we show that nonlinearity is important for their dynamics, as it allows for their spontaneous formation from an initial Gaussian wave packet. Unique properties of exciton-polaritons may lead to applications of their X-waves in long-distance signal propagation inside novel integrated optoelectronic circuits based on excitons.
\end{abstract}

\maketitle

{\it Introduction.} Sending an optical signal without distortion necessitates control over diffraction and dispersion of the traveling pulse. This can be achieved by creation of strongly confined, propagation-invariant wave packets, which behave like particles rather than waves. Perhaps the most well-known example of such a wave packet is a soliton, which is a self-localized wave existing due to the balance between nonlinearity and dispersion. It is less known that a wide class of qualitatively different solutions, called localized waves, exists and can be more usable than solitons in real-life applications~\cite{LocalizedWaves,Day2004}. In particular, they exist even in the linear regime, where there is no self-localization mechanism as in the case of solitons. The most interesting localized waves, the X-shaped optical or acoustic pulses, have been applied in numerous practical situations, including large depth of field or high-frame-rate medical imaging, optical tomography and high capacity communications~\cite{LocalizedWaves,Day2004,Lu1990,Lu1999,Wade2000}. They are nonmonochromatic, yet nondispersive superpositions of nondiffracting Bessel beams~\cite{Durnin1987} with characteristic biconical X-shape.

Recently, bright exciton-polariton solitons were observed in a semiconductor microcavity~\cite{Sich2012,Dominici2013}, and the potential of traveling polariton pulses for applications in information processing has been pointed out~\cite{Liew2008,Adrados2011,Gao2012,Ballarini2013}. The study of quantum coherent phenomena of exciton-polaritons in optical microcavities is nowadays a very active area of research~\cite{Colas2016,Kasprzak2006,Deng2010,Byrnes2014,Sedov2015}. 

Exciton-polaritons are quantum quasiparticles which are superpositions of cavity photons and excitons (electron-hole pairs) in a semiconductor~\cite{Microcavities}. Their composite nature gives rise to mixed properties characteristic to both light and matter. The strong nonlinear interactions between the excitons links the polaritons inherently with nonlinear quantum processes. The photonic part gives rise to an extremely light effective mass and allows for a direct detection of the system evolution on a picosecond timescale. 

\begin{figure}[t]
\includegraphics[height=4cm]{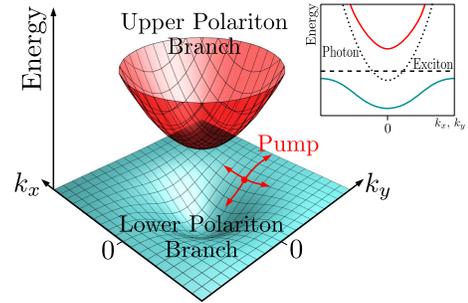}
\caption{(Color online) Dispersion relation for exciton-polaritons in a planar microcavity encompasses the upper and the lower polariton branch. The latter reveals a saddle inflection point (red) which enables creation of polaritonic X-waves. The inset compares the dispersion for photons, excitons and polaritons in one dimension.}
\label{DispersionRelation}
\end{figure}

Here, we show that the creation of confined propagation-invariant exciton-polariton wave packets does not have to rely on the soliton concept~\cite{Sich2012,Gersten1975,Silberberg1990,Liu1999,Eisenberg2001,Egorov2009,Egorov2010,Xue2014,Pinsker2014,Smirnov2014,Kulczykowski2015}. It was pointed out~\cite{Egorov2009,Egorov2010} that fully two-dimensional bright polariton solitons can exist due to the peculiar dispersion relation with positive and negative effective masses in orthogonal directions, but only in a very narrow range of pumping powers. Instead of relying on the nonlinearity, we demonstrate that in the same dispersion regime localized X-wave pulses can be created and propagate without distortion over long distances even in the low-density linear regime. There is no necessity to tune the pumping power precisely; in fact the wave packet remains nondiffractive even in the absence of pumping, as we demonstrate in direct simulations of the mean-field Gross-Pitaevskii equations. The nonlinearity, however, plays an important role in the creation of X-waves. We show that polariton nonlinear X-waves develop spontaneously from a simple Gaussian input pulse in a four-wave mixing process~\cite{Boyd2003}. We note that Bragg X-wave solutions were previously considered in a setup consisting a large number of quantum wells~\cite{Sedov2015}. In contrast, here we show that X-waves can be created in a typical sample with one or several quantum wells.

Uniqueness of the exciton-polariton X-waves is manifested by their potential for applications in exciton-based optoelectronic systems. Transistors, spin-based switches and even full low-level logic gates have been already demonstrated~\cite{Ballarini2013,Wada2004,Amo2010}. They operate at 100\kern.25em GHz--10\kern.25em THz range of frequencies, in this way filling a gap between electronic and photonic systems~\cite{Feurer2007}. These speeds of operation are suitable for processing information at rates exceeding terabits per second, necessary in the contemporary world of information~\cite{vanUden2014}. However, up to date, dispersion effects played a detrimental role in the signal guidance within polaritonic circuits~\cite{Ward2005}. Localized X-waves propagate over long distances, transferring information within the circuits with high speed and low power requirements. This contributes to making exciton-polaritons a promising platform, which may overcome recent slowdown of development in silicon-based chip architecture~\cite{Hilbert2011}, attributed to heat dissipation and resonance barriers~\cite{Ward2005}.

{\it Exciton-polaritons.} Typically, exciton-polaritons are created in planar microcavities, consisting of one or several quantum wells placed between two Bragg reflectors, and pumped by a laser beam. Excitons are formed in wells  located at the antinodes of the photonic mode. Just like photons, they are strongly confined in the direction along the cavity axis, but are free to move in the transversal x-y plain. The cavity introduces strong coupling between excitons and photons, leading to formation of new bosonic quasiparticles -- the exciton-polaritons -- which are superpositions of the photonic and excitonic component with wave functions $\psi_c$  and $\psi_x$, respectively. Their dispersion relation, shown in Fig.~\ref{DispersionRelation}, results from crossing of dispersions for photons and excitons and consists of two branches: for the lower and the higher energy solution, the lower (LP) and upper polariton, respectively. We focus exclusively on the LPs. Their dispersion reveals a saddle inflection point $k_{\text{inflection}}$, highlighted in red, which is the crucial feature enabling generation of X-waves. They possess a relatively light effective mass, determined by the photon dispersion, and a rather short lifetime, up to $100\kern.25em\mathrm{ps}$, due to the leakage of cavity photons. Thus, external pumping is necessary for creating the population of polaritons at a level necessary for obtaining the superfluid phase. We assume this is done by a strong pump with momentum $k_0$, tuned to the saddle point in the dispersion (highlighted red in Fig.~\ref{DispersionRelation}), and the LPs are created in a coherent fashion near this point. 

{\it Polaritonic X-waves.} The first sign of existence of polaritonic X-waves comes from the simplified description of LPs, in terms of a single wave function, disregarding their compound nature. In this approach the LP superfluid can be directly mapped to an atomic Bose-Einstein condensate (BEC) in a potential of an external lattice, the system where the matter X-waves are recognized~\cite{Conti2004}. Thus, the dynamics of  polaritons are well described in the mean field approximation by the generic nonlinear Schr\"odinger equation, equivalent to the Gross-Pitaevskii (GP) equation~\cite{LocalizedWaves}
\begin{equation}
  \label{eq:simple_X}
  i\hbar\dfrac{d\psi}{dt}
  + \dfrac{\hbar^2}{2m_x} \Big(\partial^2_x
  - \dfrac{m_x}{m_y}\,\partial^2_y\Big)\psi
  - \dfrac{3g}{2}|\psi|^2\psi= 0.
\end{equation}
Here $\psi$ denotes the LP wave function which corresponds to the BEC's one, $t$ is time, $\hbar$ is the Planck constant, $g$ quantifies the polariton-polariton interaction and mimics the interaction in the BEC. The quasiparticle-character of LPs is captured by a peculiar feature that their mass measured in the direction $x$ and $y$ is different: $m_x$ and $m_y$, respectively. This allows to complete the analogy to the BEC system by identifying e.g.\ $m_x$ with the atomic mass and $-m_y$ with the negative mass associated with the lattice (or vice versa). Important feature of (\ref{eq:simple_X}) is its hyperbolic form -- the signs in front of the second derivatives are opposite. This provides (\ref{eq:simple_X}) with stationary X-wave solutions.

\begin{figure}\centering
\begin{tabular}{ccc}
\null\quad a)& \null\quad b)& \null\quad c)\\
\includegraphics[height=2.5cm]{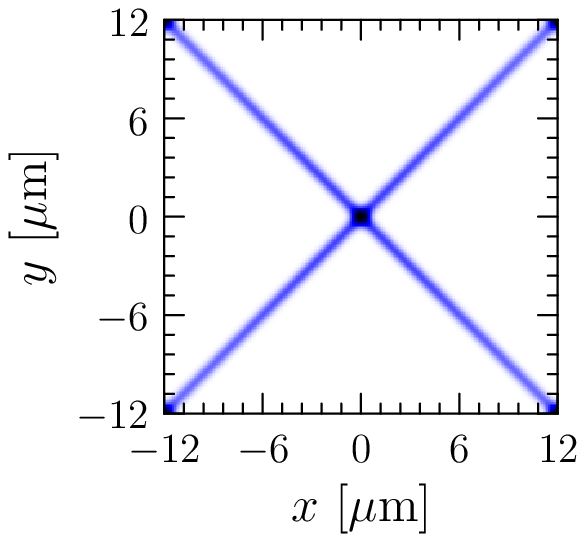}&
\includegraphics[height=2.5cm]{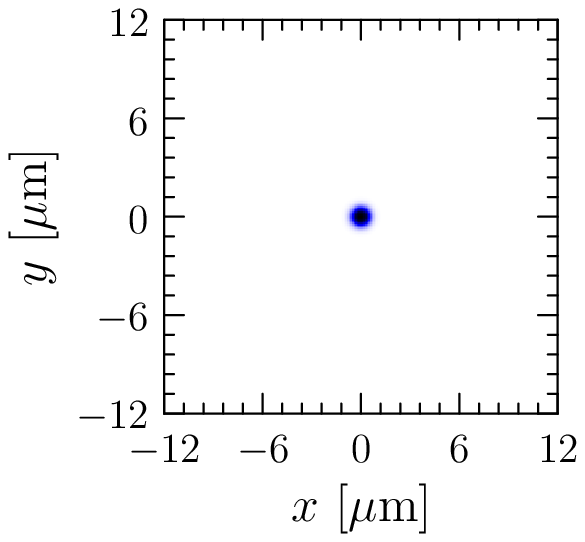}&
\includegraphics[height=2.5cm]{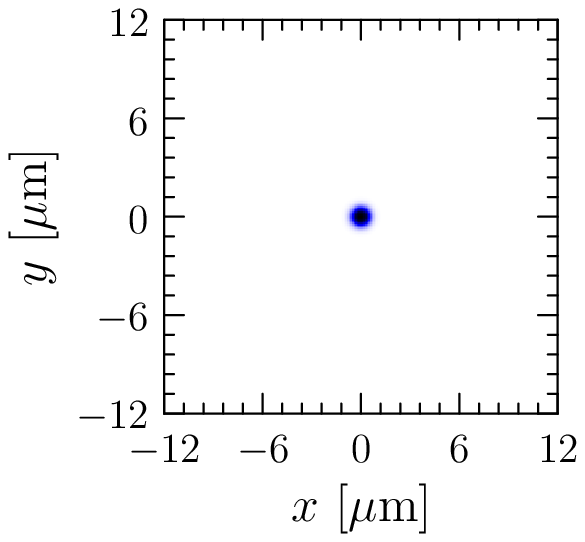}\\
\includegraphics[height=2.5cm]{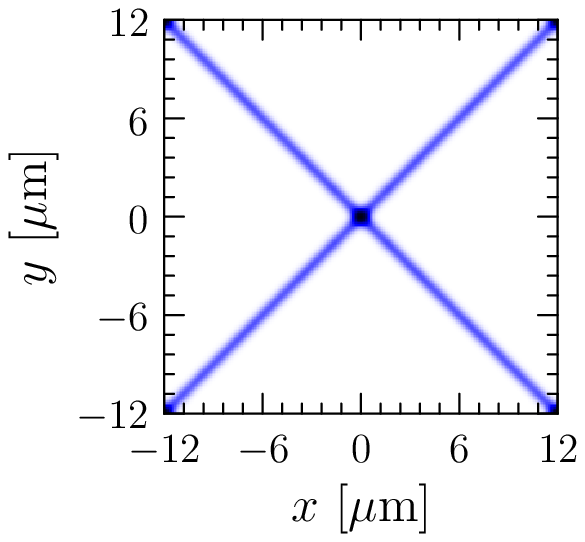}&
\includegraphics[height=2.5cm]{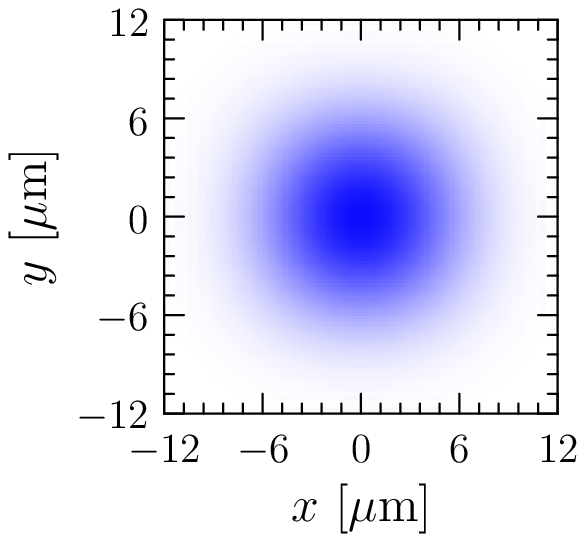}&
\includegraphics[height=2.5cm]{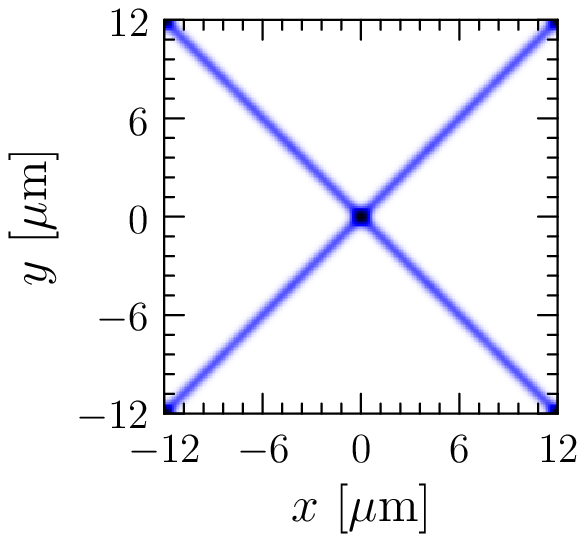}
\end{tabular}
\caption{(Color online) Evolution of $\psi(x,y)$ (upper row -- $t=0$, bottom row -- $t=2\kern.25em\mathrm{ps}$) in the generic hyperbolic model (\ref{eq:simple_X}) of a) an X-shaped stationary state, b) an initial narrow Gaussian state, the state spreads in time. c) In the presence of nonlinearity $g=5\times10^{4}\kern.25em\mathrm{\mu eV\cdot (\mu m)^2}$, an initial Gaussian state spontaneously turns into an X-shaped state. Computations were performed for $m_x=m_y=2\times10^{-5}m_e$.}
\label{fig:simple_linear_X_stationary}
\end{figure}

We will first seek for the family of stationary solutions of (\ref{eq:simple_X}) for the linear case $g=0$ in the form of plane waves $\psi(x,y) =\psi^0 e^{ik_x x + ik_y y -\omega t}$. In particular, we are interested in the family of solutions fulfilling the condition $k_y = \pm \sqrt{\tfrac{m_y}{m_x}} k_x$, i.e.\ $\omega=0$. A general stationary solution can be easily constructed using its Fourier transform $\psi(x,y) = \tfrac{1}{2\pi}\int_{\mathbb{R}^2} \tilde{\psi}(k_x,k_y) e^{ik_x x + ik_y y} dk_x dk_y$, with $\tilde{\psi}(k_x,k_y)$ fulfilling the above relation between $k_y$ and $k_x$. We take its simplest form being a convolution of an arbitrary envelope function, e.g.\ a Gaussian in $k_x$, and a characteristic function $\chi(k_x,k_y)$ on the set of $(k_x,k_y)$, equal $1$ if $\omega=0$ and $0$ otherwise, $\tilde{\psi}(k_x,k_y) \propto e^{-k_x^2} \chi(k_x,k_y)$. This state initially X-shaped both in the configuration and the Fourier space, when subjected to the evolution set by (\ref{eq:simple_X}), indeed does not change in time as it is shown in Fig.~\ref{fig:simple_linear_X_stationary}(a). This is in contrast to a Gaussian initial state which spreads quickly, see Fig.~\ref{fig:simple_linear_X_stationary}(b). In this figure the evolution takes $2\kern.25em\mathrm{ps}$ and is computed for $m_x=m_y= 2\times10^{-5}m_e$.

Strong nonlinearity in (\ref{eq:simple_X}) introduces a dramatic change in the behavior of the system, it allows creation of the X-wave spontaneously from a Gaussian initial state, see Fig.~\ref{fig:simple_linear_X_stationary}(c) computed for $g=5\times10^{4}\kern.25em\mathrm{\mu eV\cdot (\mu m)^2}$. We note that similar behavior has been observed previously in the context of nonlinear optics, which explained the spontaneous localization of pulses in a medium with normal dispersion~\cite{DiTrapani2003,Kolesik2004,Couairon2006}. High-intensity X-waves, unlike linear ones, can be formed spontaneously through a trigger mechanism of conical emission~\cite{Conti2003}. This is a great simplification in terms of the possibility of experimental observation of X-waves, since they have a rather complex structure in the Fourier space. Nevertheless, the analysis of the linear case is very instructive for describing the shape of the X-shaped stationary state  and for obtaining optimal parameters ensuring stable numerical simulations. 

Repeating the steps analogous to the ones presented above makes it is possible to find the X-wave solutions for a set of coupled GP equations describing the polaritonic superfluid~\cite{Egorov2009,Egorov2010,Amo2009}
\begin{equation}
  \label{eq:full_X}
  \kern-0.5em\left\{\kern-0.25em\begin{aligned}[c]
  i\hbar\dfrac{d}{dt}\psi_c - \Omega_R\psi_x
  - \Big(\dfrac{\hbar\gamma_c}{2i} - \dfrac{\hbar^2}{2m_c}\partial^2_x
  - \dfrac{\hbar^2}{2m_c}\partial^2_y\Big) \psi_c ={}& 0,\\
  i\hbar\dfrac{d}{dt}\psi_x - \Omega_R\psi_c - \big(\dfrac{\hbar\gamma_x}{2i}+g\lvert\psi_x\rvert^2\big)\psi_x={}& 0,
  \end{aligned}\kern-1em\right.
\end{equation}
where $m_c$ is the effective mass of polaritons, $\Omega_R$ is the Rabi frequency coupling the excitonic and photonic modes, $\gamma_c$ and $\gamma_x$ are decay rates for the photons and excitons, respectively. Now the task is more challenging since the stationary X-wave solution for (\ref{eq:full_X}) does not exist at $k=0$. This is because the signs in front of the derivatives in $x$ and $y$ directions are the same, which reflects the fact that in a microcavity there is no distinguished direction for the polaritons. Flipping one of the signs is necessary for the existence of the X-wave.  Therefore, we will construct an initial state past the inflection point, for which second derivatives have opposite signs. To this end, we will first look for analytical plane wave solutions of (\ref{eq:full_X}) for the ideal case ($\gamma_x=\gamma_c=0$) and in the non-interacting (linear) regime $g=0$. We will assume that the waves are moving with a constant velocity $v$ in the $y$ direction, $\psi_{x,c}(x,y,t)=\phi_{x,c}(x,y-vt)e^{-i\mu t/\hbar}$. Such solutions fulfill the relation  $\tfrac{\partial\psi_{x,c}}{\partial t} = - v \tfrac{\partial\psi_{x,c}}{\partial y} -i\tfrac{\mu}{\hbar}\psi_{x,c}$, which in Fourier space translates to 
\begin{equation}
  \omega = v k_y +\tfrac{\mu}{\hbar}.
\label{eq:cond_f}
\end{equation}
Some specific values of $\omega$ and $\mu$ obeying (\ref{eq:cond_f}) provide a solution that has an X shape and is quasi-localized in space. We obtain them from the dispersion for LPs, which results from the form of the plane wave solutions $\psi_{x,c}(x,y,t)=\psi_{x,c}^0\exp(ik_xx+ik_yy-i\omega t)$ and the GP equations (\ref{eq:full_X})
\begin{equation}
\label{dispersion_LP}
\hbar\omega = \tfrac{1}{2} \left(\epsilon_k-\sqrt{\epsilon_k^2+4\Omega_R^2}\right),\quad \phi_{x} = \tfrac{\Omega_R}{\hbar \omega} \phi_{c},
\end{equation}
with $\epsilon_k=\tfrac{\hbar^2}{2m_c}(k_x^2+k_y^2)$. Note that the superposition of plane waves linked with (\ref{eq:cond_f}) amounts to a moving solution. Importantly, the dependence $\omega(k_x,k_y)$ in (\ref{dispersion_LP}) ensures $d^2\omega/dk_y^2<0$ if $k_y>k_{\text{inflection}}$ while $d^2\omega/dk_x^2>0$ when $k_x<k_{\text{inflection}}$. In our simulations we take $\Omega_R=4.4\kern.25em\mathrm{meV}$, $m_c=2\times10^{-5}m_e$ and $k_x=0$ thus, the inflection point lies at $k_{\text{inflection}}\approx 1.26\kern.25em\mathrm{\mu m^{-1}}$.

\begin{figure}\centering
\includegraphics[height=2.5cm]{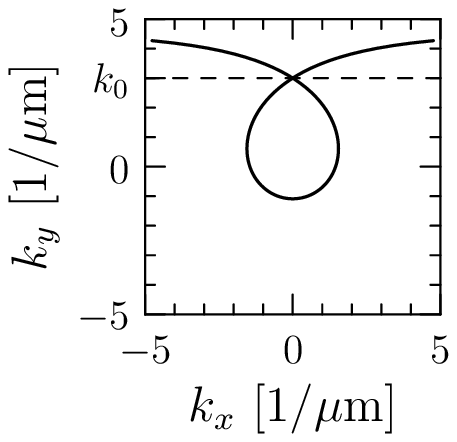}
\caption{The line depicts the points in Fourier space which fulfill the condition~(\ref{eq:full_X_parametric_fun}). Any superposition of plane waves belonging to this set is a stationary solution. In practice, only waves from the vicinity of $k_0$ are used which results in a spatially localized state. Parameters are $k_0=3\kern.25em\mathrm{\mu m^{-1}}$, $\Omega_R=4.4\kern.25em\text{meV}$, $m_c=2\times10^{-5}m_e$.}
\label{fig:full_X_solution}
\end{figure}

\begin{figure}[th]\centering
\raisebox{3cm}{a)}
\includegraphics[height=3.5cm]{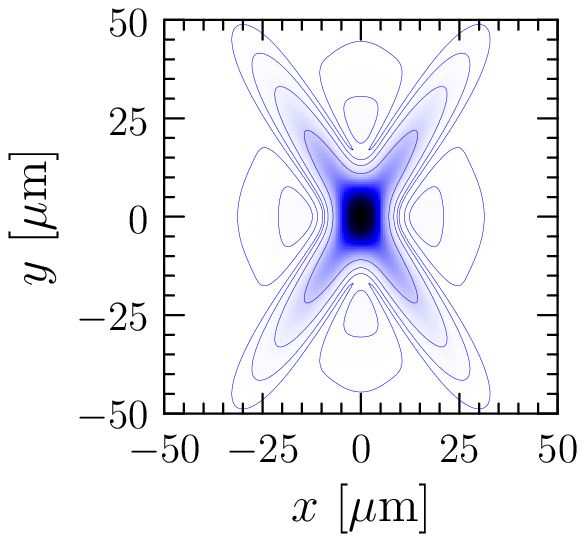}\quad
\includegraphics[height=3.5cm]{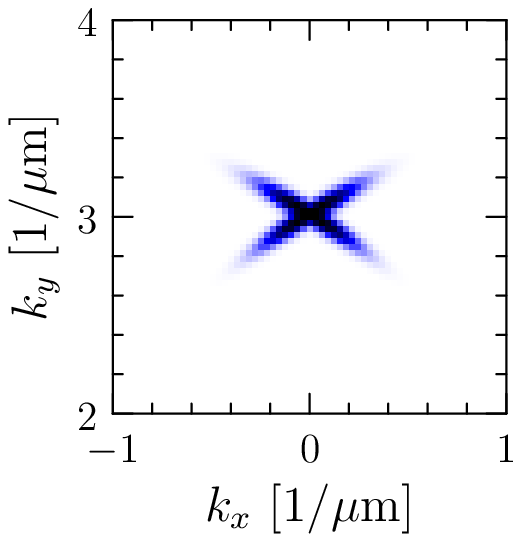}\\
\raisebox{3cm}{b)}
\includegraphics[height=3.5cm]{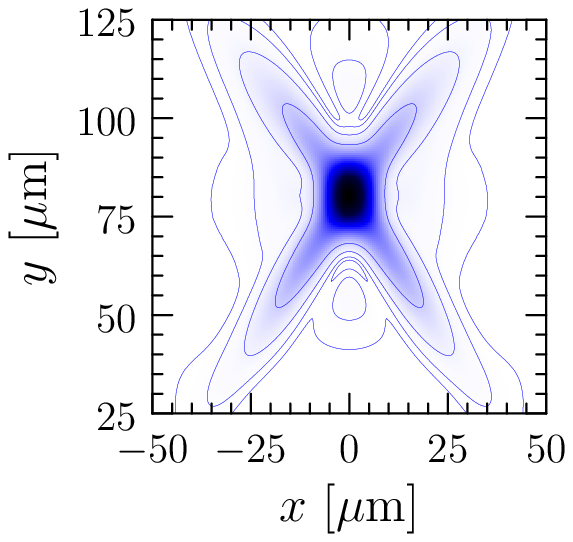}\quad
\includegraphics[height=3.5cm]{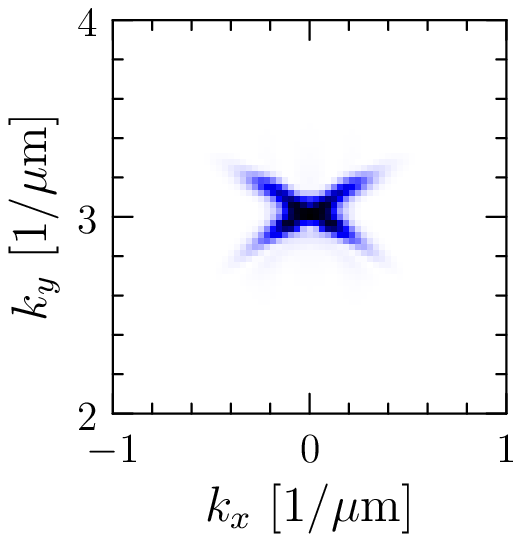}
\caption{(Color online) Evolution of (\ref{eq:full_X}) in the linear regime $g=0$ for the initial state being the X-shaped photonic stationary state (\ref{eq:Fc0}), computed for $k_0=3\kern.25em\mathrm{\mu m^{-1}}$, $\Omega_R=4.4\kern.25em\text{meV}$, $m_c=2\times10^{-5}m_e$, $\gamma_c=5\times 10^{-3}\kern.25em\mathrm{ps^{-1}}$, $\gamma_x=0.02\kern.25em\mathrm{ps^{-1}}$ and a) $t = 0$ b) $t = 100\kern.25em\text{ps}$.  Left column displays $\psi_c(x,y)$ whereas right column -- its Fourier transform, $\tilde{\psi}_c(k_x,k_y)$.  The state does not spread, but moves with a constant velocity in the $y$ direction.}
\label{fig:full_linear_X}
\end{figure}

We will now explicitly construct the X-wave solutions. We take $k_y=k_0 > k_{\text{inflection}}$ and $k_x=0$. Expansion of (\ref{dispersion_LP}) in the Taylor series around $k_y$ gives the optimal values $\hbar v = \tfrac{\epsilon_{k_0}}{k_0} - \tfrac{\epsilon_{k_0}^2}{k_0 \sqrt{\epsilon_{k_0}^2+4\Omega_R^2}}$, $\mu = \tfrac{1}{2} \left(\epsilon_{k_0} - \sqrt{\epsilon_{k_0}^2+4\Omega_R^2}\right) - \hbar v k_0$, where $\epsilon_{k_0}=\tfrac{\hbar^2}{2m_c}k_0^2$. We build the solution in its Fourier space, as before for (\ref{eq:simple_X}), but for a different characteristic function $\chi(k_x,k_y)$, which now results from (\ref{eq:cond_f}) and (\ref{dispersion_LP}). We express these conditions, linking $k_x$ and $k_y$, in the polar coordinates $(r, \theta)$
\begin{align}
\label{eq:full_X_parametric_fun}
k_x &=r(\theta)\cos\theta, \quad k_y=k_0+r(\theta)\sin\theta, 
\end{align}
with $r(\theta) = \tfrac{1}{2a_0k_0\sin\theta} - (1 + a_0k_0^2)k_0\sin\theta$, $a_0 = \tfrac{\hbar^2}{2m_c\sqrt{\epsilon_{k_0}^2+4\Omega_R^2}}$ and $\epsilon_{k_0}=\tfrac{\hbar^2}{2m_c}k_0^2$. The Fourier transforms for the photonic and excitonic wave functions equal to
\begin{align}
\label{eq:Fc0}
  \tilde{\psi}_c(k_x, k_y) &{}\propto e^{-(k_y-k_0)^2-k_x^2} \chi(k_x,k_y), \\
\tilde{\psi}_x(k_x, k_y) &{}\propto
  \tfrac{2\Omega_R}{
    \epsilon_k - \sqrt{\epsilon_k^2+4\Omega_R^2}}\tilde{\psi}_c(k_x, k_y), 
  \label{eq:Fx0}
\end{align}
where $\chi(k_x,k_y)=1$ if $(k_x,k_y)$ fulfill (\ref{eq:full_X_parametric_fun}) and $\chi(k_x,k_y)=0$ otherwise. Fig.~\ref{fig:full_X_solution} depicts $\tilde{\psi}_c(k_x,k_y)$ computed for exemplary values of $k_0=3\kern.25em\mathrm{\mu m^{-1}}$, $\Omega_R=4.4\kern.25em\mathrm{meV}$, and $m_c=2\times10^{-5}m_e$.

Having formulated the condition for the initial X-wave semi-stationary solution, Eqs.~(\ref{eq:Fc0})-(\ref{eq:Fx0}), we performed numerical simulations of the X-wave evolution using Eq.~(\ref{eq:full_X}) for the realistic system parameters, including decay, but without nonlinearity. Solutions of Eq.~(\ref{eq:full_X}) were found numerically by direct integration using the Runge--Kutta method of the fourth order. Fig.~\ref{fig:full_linear_X}(a) shows time evolution of an X-wave packet, computed for $k_0=3\kern.25em\mathrm{\mu m^{-1}}$, $\Omega_R=4.4\kern.25em\mathrm{meV}$, $m_c=2\times10^{-5}m_e$, $\gamma_c=5\times 10^{-3}\kern.25em\mathrm{ps^{-1}}$, $\gamma_x=0.02\kern.25em\mathrm{ps^{-1}}$, $g=0$, within time interval $100\kern.25em\mathrm{ps}$. These parameters correspond to state-of-the-art samples reported in literature, see e.g.~\cite{Sun2015}. The X-wave moves with a constant velocity $v$ covering the distance of around $75\kern.25em\mathrm{\mu m}$.  Any significant change in the shape of the X-wave is notable only for very late times of evolution, e.g.\ in Fig.~\ref{fig:full_linear_X}b this is $100\kern.25em\mathrm{ps}$, which by far exceeds the average lifetime of polaritons.  Since the excitonic wave function (\ref{eq:Fx0}) is just a rescaled photonic one, its evolution looks qualitatively the same.

\begin{figure}\centering
\includegraphics[height=3.5cm]{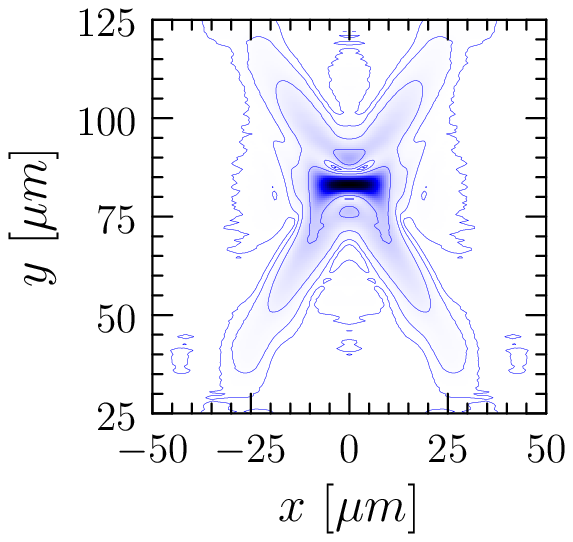}\quad
\includegraphics[height=3.5cm]{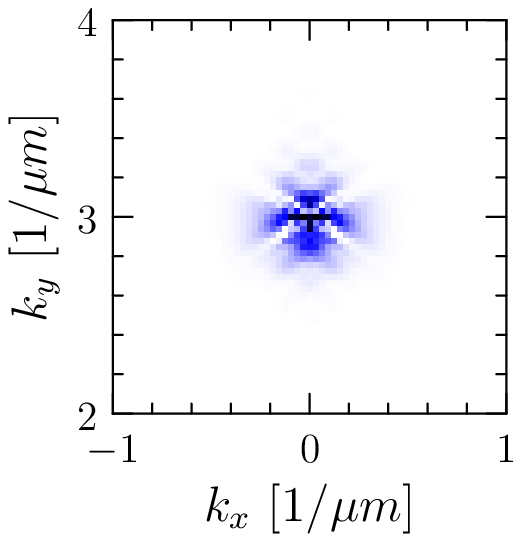}\\
\caption{(Color online) As in the previous figure, but with a Gaussian initial state and nonlinearity $g=10^{4}\kern.25em\mathrm{\mu eV\cdot (\mu m)^2}$, displayed at time $t=100\kern.25em\mathrm{ps}$. The Gaussian state spontaneously turns into an irregular X-wave. Left column shows $\psi_c(x,y)$ while right column its Fourier transform. Here $k_0=3\kern.25em\mathrm{\mu m^{-1}}$, $\Omega_R=4.4\kern.25em\text{meV}$, $m_c=2\times10^{-5}m_e$, $\gamma_c=5\times 10^{-3}\kern.25em\mathrm{ps^{-1}}$, $\gamma_x=0.02\kern.25em\mathrm{ps^{-1}}$}.
\label{fig:full_nonlinear_X}
\end{figure}

In the presence of a nonlinearity, the quasi-stationary solution of (\ref{eq:full_X}) is very similar to (\ref{eq:Fc0})-(\ref{eq:Fx0}) and the position of the inflection point does not change. However, this solution behaves as an attractor and together with the nonlinearity allows for an X-shaped wave packet to be created spontaneously from a Gaussian initial state. The phase matching condition for the four-wave mixing process in energy-momentum space corresponds exactly to the X-wave shape of our solution, as follows from our mathematical construction~\cite{Boyd2003}.

Fig.~\ref{fig:full_nonlinear_X} depicts the evolution within $t=100\kern.25em\mathrm{ps}$ evaluated for a Gaussian initial state, $k_0=3\kern.25em\mathrm{\mu m^{-1}}$, $\Omega_R=4.4\kern.25em\text{meV}$, $m_c=2\times10^{-5}m_e$, $\gamma_c=5\times 10^{-3}\kern.25em\mathrm{ps^{-1}}$, $\gamma_x=0.02\kern.25em\mathrm{ps^{-1}}$ and nonlinearity $g=10^{4}\kern.25em\mathrm{\mu eV\cdot (\mu m)^2}$. Shortly after the beginning of the evolution the characteristic tails of the X-wave develop. Note that the shape of the developed wave packet is not completely the same as in the case of Fig.~\ref{fig:full_linear_X}, but some irregularity is present both in real and Fourier space. This is in analogy to the observations made in optical Kerr media~\cite{DiTrapani2003,Kolesik2004,Couairon2006}. X-shaped quasi-stationary wave packets appeared spontaneously in the normal dispersion regime, but their internal dynamics was rather complex. 

\begin{figure}[th]\centering
\includegraphics[height=3.5cm]{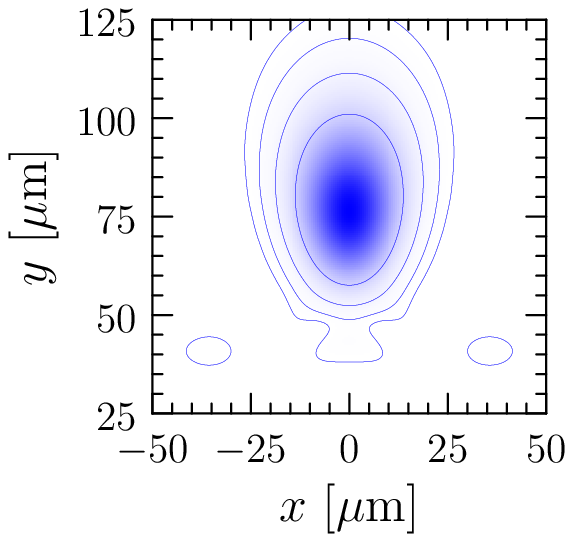}\quad
\includegraphics[height=3.5cm]{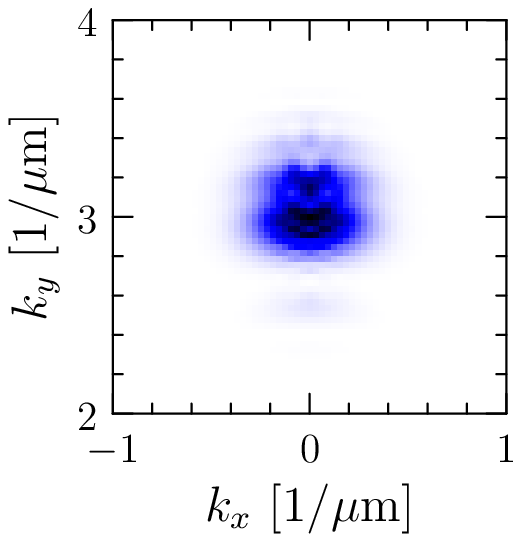}
\caption{(Color online) As in the previous figure, but in the linear regime $g=0$, shown at time $t = 100\kern.25em\mathrm{ps}$. The state decays quickly, unlike the nonlinear solution depicted in Fig.~\ref{fig:full_nonlinear_X}.}
\label{fig:full_linear_X_Gaussian}
\end{figure}

To illustrate the strength of the X-wave localization induced by the nonlinearity, we show the result of analogous simulations starting from a Gaussian initial state, but in the linear regime with $g=0$ (Fig.~\ref{fig:full_linear_X_Gaussian}). In this case, due to the dispersion, the initial Gaussian state quickly spreads and decays anisotropically. Note that the distance traveled by the wave packet is similar to that in the nonlinear X-wave case, as this is dictated by the group velocity at the dispersion point which seeded by the excitation pulse. For the same reason, the X-wave cannot be turned into the Gaussian state since it is not a quasi-stationary solution.

{\it Conclusions. } We have demonstrated the possibility to create localized X-wave solutions in resonantly excited exciton-polariton superfluids. We have constructed stationary X-wave wave packets by appropriate superposition of plane waves and demonstrated their undistorted movement over large distances. We have shown that X-waves can be created simply with a Gaussian-shaped initial pulse in the case of interacting superfluid. Contrary to bright solitons, X-waves can propagate even in the noninteracting or linear regime, hence they are more robust against the inherent polariton decay. Additionally, creation and propagation of X-waves in an exciton-polaritonic condensate does not require constant laser pumping. These properties allow X-waves to be used as transmission channels between stations without distortion, in this way contributing to the development of spin-based integrated circuits.

MS, OV, AB were supported by the EU 7FP Marie Curie Career Integration Grant No. 322150 ``QCAT'', NCN grant No. 2012/04/M/ST2/00789, MNiSW co-financed international project No. 2586/7.PR/2012/2 and MNiSW Iuventus Plus project No. IP 2014 044873.  MM acknowledges support from the National Science Center grant DEC-2011/01/D/ST3/00482. Numerical computations were carried out at the Academic Computer Centers: Cyfronet AGH in Cracow (Zeus cluster) and CI TASK in Gdansk (Galera cluster).


\begin{thebibliography}{99}

\bibitem{LocalizedWaves} H. E. Hern\'andez-Figureoa, M. Zamboni-Rached, and E. Recami (eds.), \textit{Localized Waves}, Wiley Series in Microvawe and Optical Engineering (Wiley-Interscience, 2007).

\bibitem{Day2004} Ch. Day, Phys. Today \textbf{57}, 10, 25 (2004). 

\bibitem{Lu1990} J.-y. Lu and J. F. Greenleaf, IEEE T. Ultrason. Ferr. \textbf{37}, 438, (1990). 

\bibitem{Lu1999} J.-y. Lu and S. He, Opt. Commun. \textbf{161}, 187 (1999). 

\bibitem{Wade2000} G. Wade, Ultrasonics, \textbf{38}, 1 (2000). 

\bibitem{Durnin1987} J. Durnin,  J. J. Miceli, and  J. H. Eberly, Phys. Rev. Lett. \textbf{58}, 1499 (1987). 

\bibitem{Sich2012} M. Sich, D. N. Krizhanovskii, M. S. Skolnick, A. V.Gorbach, R. Hartley, D. V. Skryabin, E. A. Cerda-M{\'e}ndez, K. Biermann, R. Hey, and P. V. Santos, Nature Photon. \textbf{6}, 50 (2012). 

\bibitem{Dominici2013} L. Dominici, M. Petrov, M. Matuszewski, D. Ballarini, M. De Giorgi, D. Colas, E. Cancellieri, B. Silva Fernandez, A. Bramati, G. Gigli, A. Kavokin, F. Laussy, and D. Sanvitto, arXiv:1309.3083. 

\bibitem{Liew2008} T. C. H. Liew, A. V. Kavokin, and I. A. Shelykh, Phys. Rev. Lett. \textbf{101}, 016402 (2008). 

\bibitem{Adrados2011} C. Adrados, T. C. H. Liew, A. Amo, M. D. Mart{\'{\i}}n, D. Sanvitto, C. Ant{\'o}n, E. Giacobino, A. Kavokin, A. Bramati, and L. Vi{\~n}a, Phys. Rev. Lett. \textbf{107}, 146402 (2011). 

\bibitem{Gao2012} T. Gao, P. S. Eldridge, T. C. H. Liew, S. I. Tsintzos, G. Stavrinidis, G. Deligeorgis, Z. Hatzopoulos, and P. G. Savvidis, Phys. Rev. B  
\textbf{85}, 235102 (2012). 

\bibitem{Ballarini2013} D. Ballarini, M.~de Giorgi, E.~Cancellieri, R. Houdr{\'e}, E. Giacobino, R. Cingolani, A. Bramati, G. Gigli, and D. Sanvitto, Nat. Commun. \textbf{4}, 1778 (2013). 

\bibitem{Colas2016} D. Colas and F. P. Laussy, Phys. Rev. Lett. \textbf{116}, 026401 (2016). 

\bibitem{Kasprzak2006} J. Kasprzak, M. Richard, S. Kundermann, A. Baas, P. Jeambrun, J. M. J. Keeling, F. M. Marchetti, M. H. Szyma\'nska, R. Andr\'e, J. L. Staehli, V. Savona, P. B. Littlewood, B. Deveaud, and Le Si Dang, Nature \textbf{443}, 409 (2006). 

\bibitem{Deng2010} H. Deng, H. Haug, and Y. Yamamoto, Rev. Mod. Phys. \textbf{82}, 1489 (2010). 

\bibitem{Byrnes2014} T. Byrnes, N. Y. Kim, and Y. Yamamoto, Nature Phys. \textbf{10}, 803 (2014). 

\bibitem{Sedov2015} E. S. Sedov, I. V. Iorsh, S. M. Arakelian, A. P. Alodjants, and A. Kavokin, Phys. Rev. Lett. \textbf{114}, 237402 (2015). 

\bibitem{Microcavities} A. V. Kavokin, J. J. Baumberg, G. Malpuech, and F. P. Laussy, {\it Microcavities} (Oxford University Press, 2007).

\bibitem{Gersten1975} J. I. Gersten and N. Tzoar, Phys. Rev. Lett. \textbf{35}, 934 (1975). 

\bibitem{Silberberg1990} Y. Silberberg, Opt. Lett. \textbf{15}, 1282 (1990). 

\bibitem{Liu1999} X. Liu, L. J. Qian, and F. W. Wise, Phys. Rev. Lett. \textbf{82}, 4631 (1999). 

\bibitem{Eisenberg2001} H. S. Eisenberg, R. Morandotti, Y. Silberberg, S. Bar-Ad, D. Ross, and J. S. Aitchison, Phys. Rev. Lett. \textbf{87}, 043902 (2001). 

\bibitem{Egorov2009} O. A. Egorov, D. V. Skryabin, A. V. Yulin, and F. Lederer, Phys. Rev. Lett. \textbf{102}, 153904 (2009). 

\bibitem{Egorov2010} O. A. Egorov, A. V. Gorbach, F. Lederer, and D. V. Skryabin, Phys. Rev. Lett. \textbf{105}, 073903 (2010). 

\bibitem{Xue2014} Y. Xue and M. Matuszewski, Phys. Rev. Lett. \textbf{112}, 216401 (2014). 

\bibitem{Pinsker2014} F. Pinsker and H. Flayac, Phys. Rev. Lett. \textbf{112}, 140405 (2014). 

\bibitem{Smirnov2014} L. A. Smirnov, D. A. Smirnova, E. A. Ostrovskaya, and Y. S. Kivshar, Phys. Rev. B \textbf{89}, 235310 (2014). 

\bibitem{Kulczykowski2015} M. Kulczykowski, N. Bobrovska, and M. Matuszewski, Phys. Rev. B \textbf{91}, 245310 (2015). 

\bibitem{Boyd2003} R. Boyd, \textit{Nonlinear Optics}, (2nd ed., Academic Press, 2003).

\bibitem{Wada2004} O. Wada, New J. Phys. \textbf{6}, 183 (2004). 

\bibitem{Amo2010} A. Amo, T. C. H. Liew, C. Adrados, R. Houdr\'e, E. Giacobino, A. V. Kavokin and A. Bramati, Nature Photon. \textbf{4}, 361 (2010). 

\bibitem{Feurer2007} T. Feurer, N. S. Stoyanov, D. W. Ward, J. C. Vaughan, E. R. Statz, and K. A. Nelson, Annu. Rev. Mater. Res. \textbf{37}, 317 (2007). 

\bibitem{vanUden2014} R. G. H. van Uden, R. Amezcua Correa, E. Antonio Lopez, F. M. Huijskens, C. Xia, G. Li, A. Sch\"ulzgen, H. de Waardt, A. M. J. Koonen, and C. M. Okonkwo, Nat. Photon. \textbf{8}, 865 (2014). 

\bibitem{Ward2005} D. W. Ward, \textit{Polaritonics: An intermediate regime between electronics and photonics,} (Doctoral dissertation, Massachusetts Institute of Technology, 2005).

\bibitem{Hilbert2011} M. Hilbert, P. L\'opez, Science \textbf{332}, 60 (2011). 

\bibitem{Conti2004} C. Conti and S. Trillo, Phys. Rev. Lett. \textbf{92}, 120404 (2004). 

\bibitem{DiTrapani2003} P. Di Trapani, G. Valiulis, A. Piskarskas, O. Jedrkiewicz, J. Trull, C. Conti, and S. Trillo, Phys. Rev. Lett. \textbf{91}, 093904 (2003). 

\bibitem{Kolesik2004} M. Kolesik, E. M. Wright, and J. V. Moloney, Phys. Rev. Lett. \textbf{92}, 253901 (2004). 

\bibitem{Couairon2006} A. Couairon, E.~Gai{\v{z}}auskas, D. Faccio, A. Dubietis, and P. Di Trapani, Phys. Rev. E \textbf{73}, 016608 (2006). 

\bibitem{Conti2003} C. Conti, S. Trillo, P. Di Trapani, G. Valiulis, A. Piskarskas, O. Jedrkiewicz, and J. Trull, Phys. Rev. Lett. \textbf{90}, 170406 (2003). 

\bibitem{Amo2009} A. Amo, D. Sanvitto, F. P. Laussy, D. Ballarini, E. del Valle, M. D. Martin, A. Lema{\^{\i}}tre, J. Bloch, D. N. Krizhanovskii, M. S. Skolnick, C. Tejedor, and L. Vi{\~{n}}a, Nature \textbf{457}, 291 (2009). 

\bibitem{Sun2015} Y. Sun, Y. Yoon, M. Steger, G. Liu, L. N. Pfeiffer, K. West, D. W. Snoke and K. A. Nelson, arXiv:1508.06698.

\end{thebibliography}
\end{document}